\documentstyle[aps,manuscript]{revtex}

\begin{document}

\title {Generalized strategies in the Minority Game}
\author{M. Hart, P. Jefferies and N.F. Johnson}
\address {Physics Department, Oxford University,
Oxford, OX1 3PU, U.K.}
\author{P.M. Hui}
\address {Department of Physics, The Chinese
University of Hong Kong, Shatin, \\
New Territories, Hong Kong}

\maketitle

\begin{abstract}
We show analytically how the fluctuations (i.e. standard deviation $\sigma$)
in the Minority Game (MG) can be made to decrease below the random coin-toss
limit if the agents use more general behavioral strategies.  This
suppression of $\sigma$ results from a cancellation between the
actions of a crowd,  in which agents act collectively and make the same
decision,  and an anticrowd in which agents act collectively by making the
opposite decision to the crowd. 

\end{abstract}
\bigskip


\newpage

The Minority Game (MG) of
Challet and Zhang\cite{challet,challet2,savit} offers a simple paradigm
for complex, adaptive systems. The MG comprises an odd number
$N$ of agents, each with $s$ strategies and a memory size
$m$, who repeatedly compete to be in the
minority. In the basic MG, where agents always use their highest scoring
strategy, the size of the fluctuations (i.e. standard deviation $\sigma$)
falls below the random, coin-toss limit as $m$ varies \cite{savit}. Cavagna
{\em
et al} \cite{sherrington} considered a fascinating generalization of the
basic
MG, the `Thermal Minority Game' (TMG), whereby  agents choose between
their strategies using an exponential probability weighting.  
Their numerical simulations demonstrated that
$\sigma$ could be pushed {\em below} the random coin-toss limit just by
altering
the relative probability weighting, which plays the role of 
temperature, of the strategies\cite{sherrington}. 
As pointed out by Marsili {\em et al} \cite{marsili},
such a
probabilistic strategy weighting has a tradition in economics and encodes a
particular behavioral model.

In this brief note, we show analytically that such a reduction in the
standard
deviation $\sigma$ below the random, coin-toss limit can be understood in
terms of crowd effects\cite{crowd,us}. 
In particular, such generalized
strategy rules tend to {\em increase} 
the cancellation between the actions
of a crowd of like-minded agents, and its anti-correlated partner
(anticrowd). Our
theoretical approach builds on a recently proposed explanation of the basic
MG in
terms of such crowd effects \cite{us}.

The MG\cite{challet} comprises an
odd number of agents $N$ who choose repeatedly between
option 0 (e.g. buy) and option 1 (e.g. sell).
The winners are those in the
minority group, e.g. sellers win if there is an excess of buyers. The
outcome at
each timestep represents the winning decision, 0 or 1.
A common bit-string of the $m$ most recent
outcomes is made available to the agents at each timestep.
The agents
randomly pick $s$ strategies at the beginning of the game, with repetitions
allowed, from the pool of all possible strategies.  
After each turn, the agent assigns one (virtual) point to each of
his
strategies which would have
predicted the correct outcome. In the basic MG,
each agent uses the most successful strategy in his possession, i.e. the one
with
the  most virtual points. Because of crowd effects \cite{us}, $\sigma$ is
large
for small $m$. Here we are interested in investigating analytically the
reason why
the large
$\sigma$ in this `crowded' regime (i.e. small $m$) gets reduced below the
random limit when the strategy-picking rule is generalized, such as in Ref.
\cite{sherrington}. We hence follow the analytic crowd-anticrowd approach of
Ref.
\cite{us} for small $m$ and
$s=2$.

Consider any two strategies $r$ and $r^*$ within the list of $2^{m+1}$
strategies in the reduced strategy space \cite{challet,us}. At any moment in
the
game, the strategies can be ranked according to their virtual points, $r=1,2
\dots 2^{m+1}$ where $r=1$ is the best strategy, $r=2$ is second best, etc.
Note
that in the small $m$ regime of interest, the virtual-point strategy ranking
and popularity ranking for strategies can be taken to be identical to a good
approximation \cite{us}. Let
$p(r,r^*|r^*\geq r)$ be the probability that a given agent picks $r$ and
$r^*$,
where $r^*\geq r$ (i.e. $r$ is the best among his $s=2$
strategies). In contrast,
let
$p(r,r^*|r^*\leq r)$ be the probability that a given agent picks $r$ and
$r^*$, where $r^*\leq r$ (i.e. $r$ is the worst among
his
$s=2$ strategies). Let
$\theta$ be
the probability that the agent uses the worst of his $s=2$ strategies, while
$1-\theta$ is the probability that he uses the best. 
The
probability that the agent plays
$r$ is given by
\begin{eqnarray}
p_r & = & \sum_{r^*=1}^{2^{m+1}} [\ \theta\  p(r,r^*|r^*\leq r) + \ (1-\theta)
\ p(r,r^*|r^*\geq r)]\nonumber \\
& = & \ (1-\theta)\  p_+(r) + \ \theta\
p_-(r) + 2^{-2(m+1)}\ \theta
\end{eqnarray}
where $p_+(r)$ is the probability that the agent has picked $r$ {\em and}
that $r$
is the agent's best (or equal best) strategy; $p_-(r)$ is the probability that
the agent has picked
$r$ {\em and} that $r$ is the agent's worst strategy. It is straightforward
to
show that
\begin{equation}
p_+(r) = \bigg(\bigg[1 - \frac{(r-1)}{2^{m+1}}\bigg]^2 -
\bigg[1-\frac{r}{2^{m+1}}\bigg]^2\bigg)\ \ .
\end{equation}
Note that 
$p_+(r) + p_-(r) = p(r)$ where 
\begin{equation}
p(r) = 2^{-m} ( 1 - 2^{-(m+2)})
\end{equation}
is the probability that the agent holds strategy
$r$ after his $s=2$ picks, with no condition on whether it is best 
or worst.  
An expression for $p_-(r)$ follows from Eqs. (2) and (3).  
The basic MG\cite{challet} thus corresponds to the case $\theta = 0$. 

The TMG is a generalized version of the basic MG in which 
each agent is equipped at each timestep with his own
(biased) coin characterised by exponential probability weightings
\cite{sherrington}. An agent 
then flips this coin to decide which strategy to use
\cite{sherrington}. 
To relate the present analysis to the TMG in Ref.\cite{sherrington},  
we consider $0\leq\theta \leq 1/2$: 
$\theta=0$ roughly corresponds to
`temperature' $T=0$ \cite{sherrington}  while
$\theta\rightarrow 1/2$ roughly corresponds to $T\rightarrow \infty$,  
although we note that this correspondence is not precise (see later
discussion).  Here, it is shown 
that we can capture the essence of such
generalized strategy play without 
having to include the detailed stochastics for
each agent. Consider the mean number of agents playing strategy $r$ which is
given by
\begin{equation}
n_r = N p_{r} = N\ (1-2\theta)\  p_+(r) + N\ \theta \ p(r) +
2^{-2(m+1)}\ N\ \theta \ \ . 
%2^{-m} (1-2^{-(m+2)})\ \ .
\end{equation}
If ${n}_r$ agents use the same strategy $r$, then
they will act as a `crowd', i.e. they will make the same decision. If
${n}_{\bar r}$ agents simultaneously use the
strategy
$\bar r$ anticorrelated to $r$, 
they will make the   
opposite (anticorrelated) decision and will hence act as an
`anticrowd' \cite{us}. For small $m$, which is our regime of interest,
crowds are
sizeable while anticrowds tend to be small in the basic MG. 
For general values of $\theta$, the standard
deviation $\sigma_\theta$ in the 
number of agents making a particular decision, say 0, 
is given by \cite{us}
\begin{equation}
\sigma_{\theta} =
\bigg[ \frac{1}{2} \sum_{r=1}^{2^{m+1}}
\frac{1}{4}|n_r-n_{\bar r}|^2 \bigg]^{\frac{1}{2}}\ \ .
\end{equation}
Substituting Eqs. (3) and (4) 
for $r$ and ${\bar r} =2^{m+1} + 1 - r$ into Eq. (5)
yields
\begin{equation}
\sigma_\theta = |1-2\theta| \ \sigma_{\theta=0}
\end{equation}
where
\begin{eqnarray} \sigma_{\theta=0} = \frac{N}{\sqrt 8}\bigg[
\sum_{r=1}^{2^{m+1}} & \bigg[ &
\bigg[1 -
\frac{(r-1)}{2^{m+1}}\bigg]^2  -
\bigg[1-\frac{r}{2^{m+1}}\bigg]^2 \nonumber \\
& & +\bigg[1 -
\frac{(2^{m+1}-r+1)}{2^{m+1}}\bigg]^2 -
\bigg[1-\frac{(2^{m+1}-r)}{2^{m+1}}\bigg]^2
\bigg]^2
\bigg]^\frac{1}{2}
\end{eqnarray}
is the standard deviation for the basic MG (i.e. $\theta=0$).
Equation (6) explicitly shows that the standard deviation
$\sigma_\theta$ {\em decreases} as $\theta$ increases (recall $0\leq
\theta
\leq 1/2$): in other words, the standard deviation decreases as agents use
their worst strategy with increasing probability. An increase in $\theta$
leads
to a  reduction in the size of the larger crowds using high-scoring
strategies,
as well as an increase in the size of the smaller anticrowds using
lower-scoring
strategies, hence resulting in a more substantial 
cancellation effect between the crowd and the anticrowd. 
Note that Eq. (6) holds for all $N$ and $m$, and hence any value
of
$\sigma_{\theta=0}$ as long as the MG remains in the `crowded' regime 
given by 
$2^m<<N$. In Ref.
\cite{us}, we showed that $\sigma_{\theta=0}$ provides a reasonable fit to
the
numerical data for $m$ up to $m\sim 5-6$ with $N=101$. For small $m$ (e.g.
$m\leq
4$ for $N=101$)
$\sigma_{\theta=0}$ lies above the random coin-toss result of
${\sqrt N}/2$. As
$\theta$ increases, $\sigma_\theta$ will eventually drop {\em below} the
random
coin-toss result at $\theta = \theta_c$ where 
\begin{equation}
\theta_{c} = \frac{1}{2}- \frac{\sqrt{N}}{4}\frac{1}{\sigma_{\theta=0}}\ \ .
\end{equation}
Because of
the difference in the details for the strategy-mixing in the present
analytic
calculation as compared to Ref. \cite{sherrington}, the present analytic
model
cannot be expected to reproduce the same quantitative features as Ref.
\cite{sherrington} for
$\theta\rightarrow 1/2$ (i.e. the high temperature limit of Ref.
\cite{sherrington}). In particular, the present analytic model does not
include
the stochastics of the coin-toss involved at each time-step for each agent
which
ultimately leads to a finite
$\sigma$ as $T\rightarrow\infty$
\cite{sherrington}. Nor does it account for the fact that agents are
discrete entities \cite{us}: we have treated $n_r$ as a continuous  quantity
which is reasonable for small $m$ in the basic MG, but will become less good
an approximation as the fluctuations become small (i.e. $\sigma\rightarrow
0$) or $m$ becomes large \cite{us}. Despite these limitations, it is
remarkable that our simple analytic approach invoking crowd effects can
capture the main feature whereby
$\sigma$ falls below the random, coin-toss limit as $\theta$ (and hence
temperature
\cite{sherrington}) increases.  It also strengthens our belief  that many
results
of the MG can be understood using simple notions of crowd-anticrowd
interplay.

\end{document}